\documentclass[a4paper,11pt]{article}
\usepackage{pos}
\usepackage{comment}
\usepackage{relsize}
\usepackage{stackrel}
\usepackage{mathtools}

\newcommand{\R}{{\mathcal R}}
\newcommand{\nl}{{\mathnormal l}}
\newcommand\aprx{\stackrel{\mathclap{\normalfont\mbox{\footnotesize{dec.}}}}{\mathlarger{\approx}}}

\title{The electron EDM in the decoupling limit of the aligned 2HDM}

\author*[a]{Juan Manuel Dávila}
\author[a]{Anirban Karan}
\author[a, b]{Emilie Passemar}
\author[a]{Antonio Pich}
\author[c]{Luiz Vale Silva}

\affiliation[a]{\it Departament de F\'{i}sica Te\`{o}rica, Instituto de F\'{i}sica Corpuscular, Universitat de Val\`encia -- Consejo Superior de Investigaciones Cient\'{i}ficas, Parc Cient\'{i}fic, Catedr\'{a}tico Jos\'{e} Beltr\'{a}n 2, E-46980 Paterna, Valencia, Spain}

\affiliation[b]{\it Physics Department, Indiana University, Bloomington, Indiana 47405, USA}

\affiliation[c]{\it Departamento de Matem\'{a}ticas, F\'{i}sica y Ciencias Tecnol\'{o}gicas, Universidad Cardenal Herrera-CEU, CEU Universities, 46115 Alfara del Patriarca, Val\`{e}ncia, Spain}

\emailAdd{juandai@ific.uv.es}
\emailAdd{kanirban@ific.uv.es}
\emailAdd{passemar@ific.uv.es}
\emailAdd{pich@ific.uv.es}
\emailAdd{luiz.valesilva@uchceu.es}

\abstract{We discuss model-independent contributions to the electron EDM, focusing on those contributions emerging from a heavy scalar sector linearly realized. 
To provide a concrete new physics realization, we investigate the aligned 2HDM in the decoupling limit.
We point out that logarithmically enhanced contributions generated from Barr-Zee diagrams with a fermion loop are present in the aligned 2HDM, an effect encoded in the decoupling limit by effective dimension-6 operators, through the mixing of four-fermion into dipole operators.
The same large logarithms are absent in specific 2HDMs where a $\mathcal Z_2$ symmetry is enforced, which thus controls the basis of effective operators relevant for calculating new physics contributions to EDMs. In other words, the $\mathcal Z_2$ symmetry acts as a suppression mechanism.
In the aligned 2HDM these contributions are proportional to sources of CP violation that are potentially large, and absent in presence of the $\mathcal Z_2$ symmetry.
We then investigate the impact on the electron EDM of this extended set of free parameters.}

\FullConference{9th Symposium on Prospects in the Physics of Discrete Symmetries (DISCRETE2024)\\
 2–6 Dec 2024\\
Ljubljana, Slovenia\\}

\begin{document}
\maketitle

\section{Introduction}

Phenomena sensitive to Charge-Parity (CP) violation provide a powerful test of the Standard Model (SM) structure. In searching for New Physics (NP) sources of CP violation, Electric Dipole Moments (EDMs) play a crucial role, since experimental sensitivities have achieved exquisite levels, and SM contributions are largely suppressed.

A new scalar sector introduces a rich phenomenology in the context of CP violation. In particular, the Kobayashi-Maskawa picture can change substantially due to new complex phases entering from the scalar sector. Multiple contributions of the scalar sector to EDMs are possible.

The Two-Higgs Doublet Model (2HDM) \cite{Branco:2011iw,Gunion:1989we}, which encapsulates an additional scalar doublet compared to the SM, is one of the simplest extensions of the SM. The 2HDM suffers in general from tree-level Flavour-Changing Neutral Currents (FCNCs), which are tightly constrained experimentally. Generally, a discrete $\mathcal Z_2$ symmetry is imposed on the Lagrangian so that each type of right-handed fermion couples to one scalar doublet only, and hence tree-level FCNCs vanish \cite{Glashow:1976nt}. Nevertheless, the `Aligned Two-Higgs Doublet Model' (A2HDM) solves this issue with a much weaker requirement: flavour alignment of Yukawa couplings, i.e. the Yukawa interactions of the two scalar doublets have the same structure in flavour space \cite{Pich:2009sp, Pich:2010ic}. 

In the A2HDM the SM is extended with a second complex scalar doublet having hypercharge $Y=1/2$. After EW symmetry breaking, both doublets acquire complex vacuum expectation values (VEVs), but it is always possible to rotate the scalar doublets to the ``Higgs-basis'' where only the first doublet acquires a non-zero real VEV. In this basis, the scalar doublets take the following form:
\begin{equation}
\label{eq:Phi}
\Phi_1=\frac{1}{\sqrt 2}\begin{pmatrix}
\sqrt 2\;G^+\\
S_1+v+i\, G^0
\end{pmatrix}\, ,\qquad\qquad \Phi_2=\frac{1}{\sqrt 2}\begin{pmatrix}
\sqrt 2\;H^+\\
S_2+i\, S_3
\end{pmatrix}\, ,
\end{equation}
where $v=246$ GeV is the VEV of the $\Phi_1$ scalar and the components $G^\pm$ and $G^0$ act as Goldstone bosons. Thus, the scalar sector of this model is comprised of one pair of charged scalars $H^\pm$, two CP-even scalars $S_1$, $S_2$ and one CP-odd scalar (pseudoscalar) $S_3$. Respecting the SM gauge symmetries, the most general scalar potential 
takes the form:
\begin{align}
\label{eq:pot}
V=&\mu_1\,\Phi_1^\dagger \Phi_1+\mu_2\,\Phi_2^\dagger \Phi_2+ \Big[\mu_3\,\Phi_1^\dagger \Phi_2+h.c.\Big]+\frac{\lambda_1}{2}\,(\Phi_1^\dagger \Phi_1)^2+\frac{\lambda_2}{2}\,(\Phi_2^\dagger \Phi_2)^2+\lambda_3\,(\Phi_1^\dagger \Phi_1)(\Phi_2^\dagger \Phi_2)\nonumber\\
&+\lambda_4\,(\Phi_1^\dagger \Phi_2)(\Phi_2^\dagger \Phi_1)+\Big[\Big(\frac{\lambda_5}{2}\,\Phi_1^\dagger \Phi_2+\lambda_6 \,\Phi_1^\dagger \Phi_1 +\lambda_7 \,\Phi_2^\dagger \Phi_2\Big)(\Phi_1^\dagger \Phi_2)+ \mathrm{h.c.}\Big]\, ,
\end{align}
where the parameters $\lambda_{\{5,6,7\}}$ and $\mu_3$ can have complex values. Depending on the parameters of the scalar potential, the neutral scalars $S_1$, $S_2$, $S_3$ mix with each other through an orthogonal matrix $\mathcal R$ and produce the mass eigenstates $H_j\in\{H_1, H_2, H_3\}$, where $H_1=h$ is the lightest state. When some of the parameters in the scalar potential obtain complex values, CP-symmetry gets violated and hence the mass eigenstates do not possess any definite CP quantum number.

On the other hand, apart from the usual mass terms for the fermions $f\in\{u,d,l\}$, the interaction part of the Yukawa Lagrangian in this model becomes:
\begin{equation}
\label{eq:Yukawa}
-\mathcal L_Y=\sum_{j,f}\Bigg(\frac{y_f^{H_j}}{v}\Bigg)\,H_j\,\bar f M_f \mathcal{P}_R f+\frac{\sqrt 2 H^+}{v} \Big[\bar u\,\big\{\varsigma_d V M_d \mathcal{P}_R-\varsigma_u M_u^\dagger V\mathcal{P}_L\big\}\, d+\varsigma_\nl\, \bar \nu M_\nl \mathcal P_R\,\nl\Big] + \mathrm{h.c.}\, ,
\end{equation}
where $\mathcal P_{L,R}$ are chirality projection operators, $M_f$ are the diagonal fermionic mass matrices, and $V$ is the usual CKM matrix. The complex parameters $\varsigma_f$ are called the flavour alignment parameters. The Yukawa couplings of fermions with the neutral scalars $(y_f^{H_j})$ are given by:
\begin{equation}
y_{d,l}^{H_j}=\R_{j1}+ (\R_{j2}+i \R_{j3}) \varsigma_{d,l}^{}\qquad \text{and} \qquad y_{u}^{H_j}=\R_{j1}+ (\R_{j2}-i \R_{j3}) \varsigma_{u}^{*}~.
\end{equation}

\section{Contributions to the eEDM in the A2HDM}

It is a well-established fact that the CP violation arising from the CKM matrix of the SM is insufficient to explain the observed \textit{baryon asymmetry of the universe} \cite{Huet:1994jb}. We thus focus on 2HDM scenarios that can introduce new sources of CP violation. Usually, in these 2HDM scenarios new CP-violating terms arise from the scalar potential only. Nonetheless, the A2HDM can introduce new sources of CP violation in both the scalar and Yukawa sectors \cite{Pich:2009sp}. This CP violation in the Yukawa sector generates new two-loop contributions to the electron EDM (eEDM).

The eEDM $d_e$ can be defined as the coefficient of the following dimension-5 operator in the effective Lagrangian at a very low energy scale ($\mu\ll m_e$):
\begin{equation}
	\mathcal L \supset -\frac{i}{2} d_e (\mu)\, \bar \psi_e\, \sigma^{\mu \nu}\gamma^5 \, \psi_e\, F_{\mu\nu}
\end{equation}
where $\psi_e$ is the electron Dirac spinor and $F^{\mu\nu}$ is the field strength tensor of the photon. In the SM, long-distance hadronic contributions have been estimated to 
dominate the eEDM, giving $d_e^{SM} = 5.8 \times 10^{-40} e\,\text{cm}$ \cite{Yamaguchi:2020dsy}, well below the experimentally probed value at 90\% C.L. \cite{Roussy:2022cmp}:
\begin{equation}\label{eq:current_exp_value}
   |d_e^{exp}| \, < \, 4.1\times 10^{-30} \, e\,\text{cm}
\end{equation}

In the A2HDM, the eEDM starts getting contributions at one loop, but it is a well-known fact that the two-loop effect is dominant over the one-loop one for the eEDM \cite{Barr:1990vd}, since these diagrams contain a lower number of electron-scalar interaction vertices which generate a mass suppression. Depending on the number of gauge bosons connected to the electron line, the dominant two-loop diagrams contributing to the eEDM in the A2HDM can be classified in two classes: a) Barr-Zee (one gauge boson) and b) Kite (two gauge bosons). Based on the gauge boson connecting the electron line and the internal loop in the Barr-Zee diagrams, they can be divided into three categories: i) electromagnetic-current mediated ($\gamma$), ii) neutral-current mediated ($Z$) and iii) charged-current mediated ($W^\pm$), each of which are induced by three types of loops: fermion loop, charged Higgs loop and gauge boson loop. We will refer to i)-iii) as EM, NC, and CC contributions. The kite diagrams generate two types of contributions from the neutral and charged currents. Thus, the total contribution of the A2HDM to the eEDM is given by:
\begin{equation}
    d_{e}^{}= \sum_{L,X}d_{e,L}^{X}+ (d_{e,\rm{Kite}}^{CC}+d_{e,\rm{Kite}}^{NC})\, ,
\end{equation}
where $L \in \{f,\, H^\pm,\, W \}$ and $X \in $ \{CC,\, NC,\, EM\}. 

While comparing to the Complex 2HDM (C2HDM), which is a special case of the A2HDM, the eEDM in the A2HDM acquires contributions from the charged-current fermion-loop Barr-Zee diagrams, which are completely absent in the C2HDM scenario \cite{Altmannshofer:2020shb}. The complex coupling of the $\bar e \nu H^+$ vertex, which is real valued in the C2HDM, generates this contribution in the A2HDM.

\subsection{Full model contributions to the eEDM in the decoupling limit of the A2HDM}

The situation where the mass parameter of the second doublet $\Phi_2$ becomes very large compared to the VEV of $\Phi_1$, i.e. $\sqrt{\mu_2}\gg v$, is called the \textit{decoupling limit}. Such case can be thought of as the doublet $\Phi_2$ sitting at a very high energy scale and hence decoupling from the SM which is at a much lower energy scale. Using the masses of the particles as independent parameters, the condition for achieving the decoupling limit can be recast as $M_{\{H^\pm,\,H_2,\,H_3\}}\approx M \gg m_h$. 

In this work, we explicitly computed the leading logarithmically enhaced contributions in the decoupling limit of the A2HDM. Thus, we get a squared logarithm term coming from fermion-loop Barr-Zee contributions with a dependence on $\varsigma_u$:
\begin{align} \label{eqn:BZf_dec_u}
    \frac{d_{e, f}}{e} \Bigg|_{\mathlarger{\varsigma_u}} \, = \, \sum_{X} \frac{d_{e, f}^X}{e} \Bigg|_{\mathlarger{\varsigma_u}} \, \aprx \, m_e \frac{g^2}{(4 \pi)^4 v^2} (3+5t_w^2) \text{Im}(\varsigma_u^* \varsigma_l) \frac{m_t^2}{M^2} \log^2\left( \frac{M^2}{m_t^2} \right),
\end{align}
and two single logarithmic terms, one of them coming from fermion-loop Barr-Zee contributions with a dependence on $\varsigma_d$ and the other coming from the sum of the gauge boson loop Barr-Zee contributions and charged-current kite diagrams \cite{Altmannshofer:2020shb}:
\begin{align}
    \frac{d_{e, f}}{e} \Bigg|_{\mathlarger{\varsigma_d}} \, = \;\; \sum_{X} \frac{d_{e, f}^X}{e} \Bigg|_{\mathlarger{\varsigma_d}} \, &\aprx \;\; - m_e \frac{g^2}{(4 \pi)^4 v^2}\frac{t_w^2}{2} \text{Im}(\varsigma_d^* \varsigma_l) \frac{m_b^2}{M^2} \log\left( \frac{M^2}{m_t^2} \right), \\
    \sum_{X} \frac{d_{e, W}^X}{e} + \frac{d_{e,\rm{Kite}}^{CC}}{e} \, &\aprx \, m_e \frac{g^2}{(4 \pi)^4} \frac{3}{4} t_w^2 \, \frac{\text{Im} (\lambda_6^* \varsigma_l)}{M^2} \log \left( \frac{M^2}{m_W^2} \right).
    \label{eqn:single_log}
\end{align}

Our result agrees with the recent Ref.~\cite{Altmannshofer:2024mbj} where the eEDM has been discussed within the General 2HDM (G2HDM). In the following section, we will make it clear the origin of this pattern of powers of large logarithms in the decoupling limit.

\subsection{Model-independent contributions to the eEDM in the SMEFT}

In order to better understand the features of the A2HDM in the decoupling limit, we are interested in using the EFT approach. This will only be valid when the scale of NP (i.e. the masses of the heavy scalars) is much larger than the EW scale. This way, we can characterize their effects via Wilson coefficients $C_i$ accompanying effective operators $Q_i$ of dimension higher than 4:
\begin{equation}
    \mathcal{L} \, = \, \mathcal{L}_{SM} + \sum_i C_i (\mu) Q_i.
\end{equation}
Once generated, the coefficients of interest will run from the NP scale down to the EW scale, mixing with the coefficients of the EW dipole operators, from which we find the eEDM coefficient:
\begin{equation}
    d_e \, = \, - \sqrt{2} v \Big( c_w \text{Im} (C_{e B}) - s_w \text{Im} (C_{e W}) \Big).
\end{equation}

The leading contribution to the eEDM will come from dimension-6 operators from the SM Effective Field Theory (SMEFT). In order to reproduce the leading logarithmic contributions that appear in the decoupling limit of the A2HDM, we only need to take into account four independent effective non-dipole operators, that are conveniently expressed in the Warsaw basis \cite{Grzadkowski:2010es}.

First, the dimension-6 Yukawa operator $Q_{eH}$ and the four-fermion scalar operator $Q_{ledq}$:
\begin{align}
    Q_{eH}^{pr} \, &= \, (H^{\dagger}H)(\bar{l}_p e_r H),\\
    Q_{ledq}^{prmn} \, &= \, (\bar{l}_p^j e_r) (\bar{d}_m q_n^j),
\end{align}
both of which mix at the two-loop level into the EW dipole operators \cite{Panico:2018hal}, generating respectively two single logarithm contributions:
\begin{align}
    \frac{d_{e, \, eH}^{\text{SMEFT}}}{e} \, &= \, m_e \frac{g^2}{(4 \pi)^4} \frac34 t_w^2 \, \frac{\text{Im} (\lambda_6^* \, \varsigma_l)}{M^2} \log \left( \frac{M^2}{m_{EW}^2} \right), \\
    \frac{d_{e,b}^{\text{SMEFT}}}{e} \, &= \, - m_e \frac{g^2}{(4 \pi)^4 v^2} \frac{t_w^2}{2} \text{Im}(\varsigma_d^* \varsigma_l) \frac{m_b^2}{M^2} \log \left( \frac{M^2}{m_{EW}^2} \right).
    \label{eq:ledq_SMEFT}
\end{align}

And then, the four-fermion scalar operator $Q_{lequ}^{(1)}$:
\begin{equation}
    Q_{lequ}^{(1), prmn} \, = \, (\bar{l}_p^j e_r) \epsilon_{jk} (\bar{q}_m^k u_n),
\end{equation}
which mixes at the one-loop level into the four-fermion tensor operator $Q_{lequ}^{(3)}$:
\begin{equation}
    Q_{lequ}^{(3), prmn} \, = \, (\bar{l}_p^j \sigma_{\mu \nu} e_r) \epsilon_{jk} (\bar{q}_m^k \sigma^{\mu \nu} u_n),
\end{equation}
which gets mixed into the EW dipoles generating a squared logarithm contribution to the eEDM:
\begin{align}\label{eq:lequ_SMEFT}
    \frac{d_{e,t}^{\text{SMEFT}}}{e} \, = \, m_e \frac{g^2}{(4 \pi)^4 v^2} (5 t_w^2 + 3) \text{Im}(\varsigma_u^* \varsigma_l) \frac{m_t^2}{M^2} \log^2 \left( \frac{M^2}{m_{EW}^2} \right).
\end{align}
\section{Phenomenology}
We now explore how the inclusion of the novel charged-current fermion-loop Barr-Zee contributions affects the prediction of the eEDM. To that end, we fix the A2HDM parameters to the following benchmark values consistent with Ref.~\cite{Coutinho:2024zyp}:
\begin{align}\label{eq:benchmark_values}
\lambda_3 &= 0.02 \,,  &  \lambda_4 &= 0.04 \,, &  \lambda_7 &= 0.03 \,, & \text{Re}(\lambda_5) &= 0.05 \,, \nonumber \\
\text{Re}(\lambda_6) &= -0.05 \,, & \text{Im}(\lambda_6) &= 0.01 \,, & \alpha_3 &= \pi/6.
\end{align}

\begin{figure}[h]
    \centering
    \includegraphics[scale=0.21]{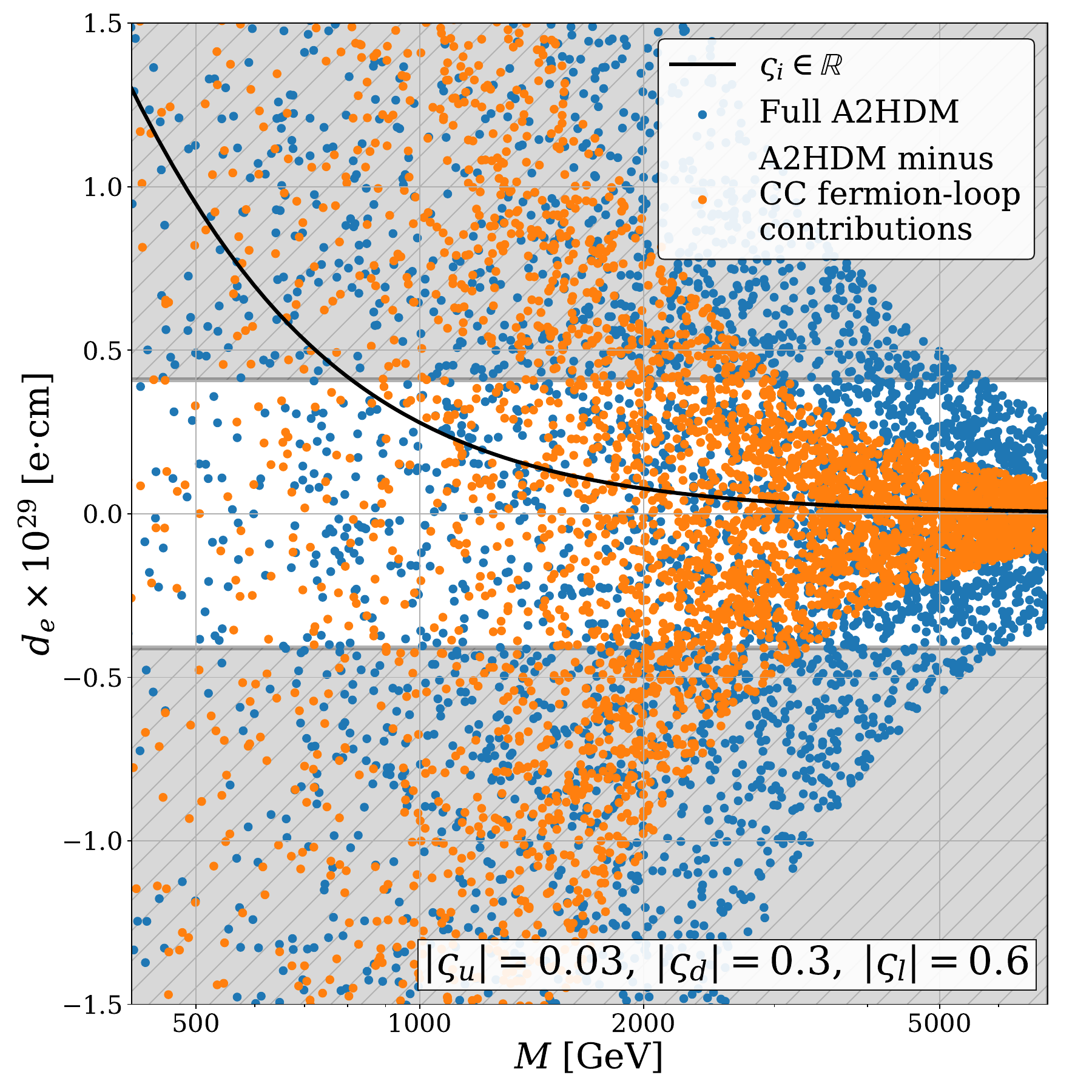}
    \hspace{8mm}
    \includegraphics[scale=0.21]{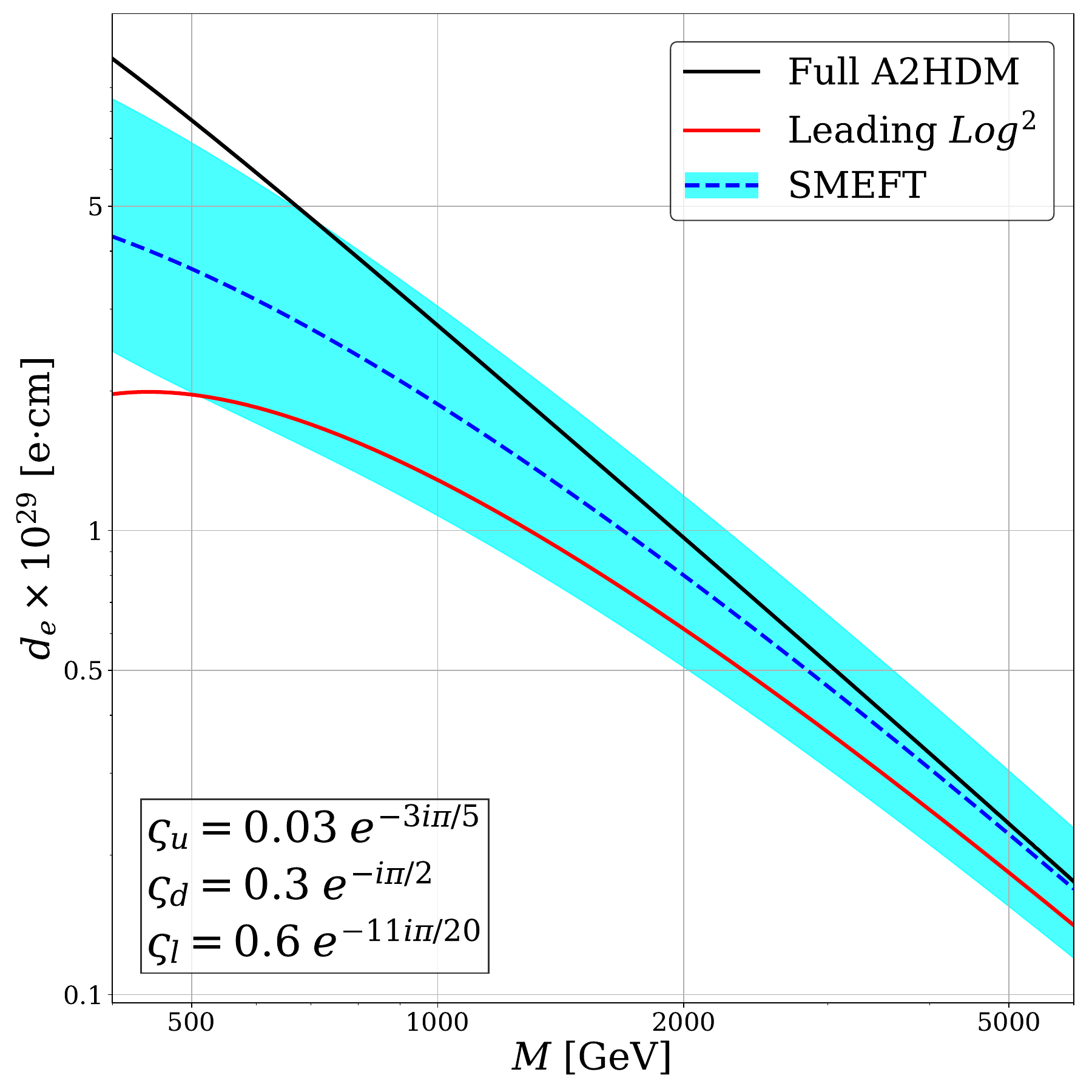}
    \caption{\textbf{Left}: Scatter plot of the eEDM in the A2HDM as a function of $M$. The blue dots correspond to the full A2HDM, while the expression at the origin of the orange dots does not include CC fermion-loop Barr-Zee contributions. The black line assumes real alignment parameters. The gray bands show the upper bound on the modulus of the eEDM. \textbf{Right}: Approximations to predictions of the eEDM in the A2HDM as a function of $M$. The black line is the full two-loop result in the A2HDM. The solid red curve is the leading squared logarithmic approximation, and the dashed blue curve includes the sub-leading logarithms from SMEFT. The shaded blue region is obtained by varying the UV scale from $M/2$ up to $2M$.}
    \label{fig:pheno}
\end{figure}

In Fig.~\ref{fig:pheno} (left panel), we show a scatter plot of the eEDM, where random values are given to the mass of the charged scalar, $M_{H^{\pm}}=M$, and the phases of the flavour alignment parameters, that is, $\varsigma_i$ with $i=u,d,l$. The blue dots are predictions from the full A2HDM, while the orange ones are computed by subtracting the new charged-current fermion-loop Barr-Zee contributions from the full result. The black line shows the prediction for a scenario with real flavour alignment parameters. Finally, we show in gray the upper bound on the modulus of the eEDM from Eq.~\eqref{eq:current_exp_value}.

In Fig.~\ref{fig:pheno} (right panel), we numerically compare various approximations to the eEDM as a function of the mass of the charged scalar, $M_{H^{\pm}}=M$. All other parameters are fixed according to the benchmark point in Eq.~\eqref{eq:benchmark_values}. The black line shows the result of the full two-loop calculation in the A2HDM. The solid red curve shows the leading squared logarithmic approximation. For the values of $M$ displayed in the plot and the mentioned benchmark point, it manages to give a rather accurate prediction for the eEDM. Finally, the dashed blue line shows the SMEFT result, and the shaded band is obtained by varying the Ultra-Violet (UV) scale at which the SMEFT logarithmic terms are evaluated between $M/2$ and $2M$.
\section{Conclusions}
We provide a discussion about the contributions stemming from the A2HDM to the eEDM. Contrary to more constrained versions of 2HDMs, the A2HDM carries extra sources of CP violation in the fermionic sector, which leads to new contributions to EDMs.

Our main focus concerns the contributions from the A2HDM in the decoupling limit. The features observed in this limit can be explained by relying on the analysis of dimension-6 operators of SMEFT. We then explicitly show that the leading logarithmic contributions are described by a few effective dimension-6 operators. Some of these contributions are absent in the more studied 2HDMs where a discrete $\mathcal Z_2$ symmetry controls the ways in which logarithmically enhanced contributions can appear, providing thus a suppression mechanism.

In our phenomenological study, we find that the contributions from charged-current fermion-loop Barr-Zee contributions can be dominant with respect to the other contributions to the eEDM. Beyond the cancellation mechanism operating when a $\mathcal{Z}_2$ symmetry is present, thus controlling the presence of some logarithmically enhanced contributions, accidental cancellation mechanisms can also operate, for instance between fermionic and non-fermionic Barr-Zee diagrams.

It would be interesting to investigate which other effects are exclusive to the A2HDM, or even more general 2HDM realizations, but that are absent in more constrained versions of 2HDMs.\\

\textbf{Acknowledgements}: We are happy to thank Wolfgang Altmannshofer, Joachim Brod, Sebastian J\"ager, Kevin Monsalvez Pozo, Verónica Sanz, Mustafa Tabet for useful discussions.
This work is supported by the Spanish Government (Agencia Estatal de Investigación MCIN/AEI/10.13039/501100011033) Grants No.  PID2020–114473GB-I00 and No. PID2023-146220NB-I00, and CEX2023-001292-S (Agencia Estatal de Investigación MCIU/AEI (Spain) under grant IFIC Centro de Excelencia Severo Ochoa).
We receive support from the Generalitat Valenciana (Spain) through the plan GenT program CIDEGENT/2021/037.
The work of JMD is also supported by Generalitat Valenciana, grant CIACIF/2023/409.
The work of EP is also supported by the U.S. National Science Foundation under
grant PHY-2310149.

\bibliographystyle{JHEP}
\bibliography{mybib}

\end{document}